\documentclass[aps,prd,reprint,groupedaddress]{revtex4-1}
\usepackage{amsmath,graphicx}

\begin{document}


\title{Polytropic spheres containing region of trapped null geodesics}
   \author{Jan Novotn\'{y}}
   \email[]{jan.novotny@fpf.slu.cz}
   \author{Jan Hlad\'{i}k}
   \email[]{jan.hladik@fpf.slu.cz}
   \author{Zden\v{e}k Stuchl\'{i}k}
   \email[]{zdenek.stuchlik@fpf.slu.cz}

\affiliation{Institute of Physics and Research Centre of Theoretical Physics and Astrophysics, Faculty of Philosophy and Science, Silesian University in Opava, Bezru\v{c}ovo n\'{a}m. 13, CZ-746\,01 Opava, Czech Republic}

\date{\today}


\begin{abstract}
We demonstrate that in the framework of standard general relativity polytropic spheres with properly fixed polytropic index $n$ and relativistic parameter $\sigma$, giving ratio of the central pressure $p_\mathrm{c}$ to the central energy density $\rho_\mathrm{c}$, can contain region of trapped null geodesics. Such trapping polytropes can exist for $n > 2.138$ and they are generally much more extended and massive than the observed neutron stars. We show that in the $n$--$\sigma$ parameter space the region of allowed trapping increase with polytropic index for interval of physical interest $2.138 < n < 4$. Space extension of the region of trapped null geodesics increases with both increasing $n$ and $\sigma > 0.677$ from the allowed region. In order to relate the trapping phenomenon to astrophysically relevant situations, we restrict validity of the polytropic configurations to their extension $r_\mathrm{extr}$ corresponding to the gravitational mass $M \sim 2M_{\odot}$ of the most massive observed neutron stars. Then for the central density $\rho_\mathrm{c} \sim 10^{15}$~g\,cm$^{-3}$ the trapped regions are outside $r_\mathrm{extr}$ for all values of $2.138 < n < 4$,  for the central density $\rho_{c} \sim 5 \times 10^{15}$~g\,cm$^{-3}$ the whole trapped regions are located inside of $r_\mathrm{extr}$ for $2.138 < n < 3.1$, while for  $\rho_\mathrm{c} \sim 10^{16}$~g\,cm$^{-3}$ the whole trapped regions are inside of $r_\mathrm{extr}$ for all values of $2.138 < n < 4$, guaranteeing astrophysically plausible trapping for all considered polytropes. The region of trapped null geodesics is located closely to the polytrope centre and could have relevant influence on cooling of such polytropes or for binding of gravitational waves in their interior.
\end{abstract}

\pacs{97.10.Cv}

\maketitle


\section{Introduction}\label{intro}
Extremely compact objects having surface $R$ located under the radius $r_\mathrm{ph}$ of photon circular geodesic of the external Schwarzschild (or some generalized spherically symmetric vacuum) spacetime are important because they have to contain a region of trapped null geodesics that could be relevant for trapping of gravitational waves \cite{1999A-GRO}, or radiated neutrinos \cite{2012SHU-NTE}. Existence of the extremely compact objects has been demonstrated in the physically implausible, but principally very interesting case of spheres with uniform distribution of energy density (but radii dependent distribution of pressure) \cite{2000S-SSS,2001SHS-NGE,2004B-ESS,2011HS-PNR}. However, the models of neutron or quark stars based on the known realistic equations of state do not allow for the existence of extremely compact objects defined in this way, as in the most extreme cases there is $R \geq 3.5M > r_\mathrm{ph} = 3M$ where $M$ denotes mass of the compact star \cite{2006LP-ESN}.

Surprisingly, recent study related to the general relativistic polytropic spheres in spacetimes with the repulsive cosmological constant demonstrates possibility to obtain relativistic polytropic spheres containing near their centre a region with trapped null geodesics \cite{2016SHN-GRP}. This is an important result as surface of such polytropes can be located above $r_\mathrm{ph}=3M$ so that we could reconsider definition of the extremely compact objects restricting attention solely to the existence of trapped null geodesics region.
Although the polytropic spheres represent some physical idealization, it is well know that they represent non-relativistic ($n=1.5$) and ultra-relativistic ($n=3$) degenerated Fermi gas that can be taken quite seriously, being interesting physically especially for the ultra-relativistic Fermi gas \cite{1983ST-BHW}.

For this reason, we study in detail the existence of general relativistic polytropes containing a region of trapped null geodesics. The role of the cosmological constant is relevant only for very extended objects with radius close to the static radius of the external spacetime \cite{1983S-MTP,1999SH-SPS,2005S-IRC} and low central density \cite{2016SHN-GRP}. It is thus clear that it will be irrelevant for our study and we can abandon the influence of the cosmological constant. In order to find the regions of trapped null geodesics we use, following the paper \cite{2016SHN-GRP}, the construction of the so-called optical geometry, related to the polytrope internal spacetime, and its embedding diagrams. This is quite efficient method as in the spherically symmetric spacetimes the turning points of the optical geometry embedding diagrams correspond to the stable and unstable photon circular geodesics implying existence of region of trapped null geodesics \cite{2000SHJ-ORG}. We relate the general discussion of the trapping polytropes to situations of direct astrophysical relevance, demonstrating their strong dependence on the central energy density and restricting validity of the polytropic state equations to regions giving masses smaller than the observational limit of neutron stars mass ($M \sim 2M_\odot$). We present detailed discussion in the case of the ultra-relativistic Fermi gas with polytropic index $n=3$.


\section{Polytrope structure equations}\label{eost}
For a spherically symmetric, static spacetime, expressed in terms of the standard Schwarzschild coordinates, the line element takes the form
\begin{equation}
    \mathrm{d}s^{2} = - \mathrm{e}^{2\Phi} c^{2} \mathrm{d}t^{2} + \mathrm{e}^{2\Psi} \mathrm{d}r^{2} + r^{2} (\mathrm{d}\theta^{2} + \sin^{2} \theta \mathrm{d}\phi^{2}).
\end{equation}
The metric has two unknown functions of the radial coordinate, $\Phi(r)$ and $\Psi(r)$. The static configuration is assumed to be a perfect fluid having the stress-energy tensor
\begin{equation}
    T^\mu_{\hphantom{\mu}\nu} = (p+\rho c^{2}) U^{\mu} U_{\nu} + p\,\delta^{\mu}_{\nu},
\end{equation}
where $U^{\mu}$ denotes the 4-velocity of the fluid. In the fluid rest-frame $\rho = \rho (r)$ represents the mass-energy density and $p = p(r)$ represents the isotropic pressure.

We assume the mass-energy density and pressure related by the polytropic equation of state
\begin{equation}
    p = K \rho^{1+1/n},
\end{equation}
where constant $n$ denotes the polytropic index. $K$ denotes a constant governed by the thermal characteristics of a given polytropic configuration by specifying the density $\rho_{\mathrm{c}}$ and pressure $p_{\mathrm{c}}$ at its center --- it is determined by the total mass and radius of the configuration, and the relativistic parameter \cite{1964T-GRP}
\begin{equation}
    \sigma  \equiv \frac{p_{\mathrm{c}}}{\rho_{\mathrm{c}} c^{2}}\  = \frac{K}{c^{2}}\rho_{\mathrm{c}}^{1/n}.
\end{equation}
For a given pressure, the density is a function of temperature. Therefore, the constant $K$ contains the temperature implicitly. The polytropic equation is a limiting form of the parametric equations of state for the completely degenerate gas at zero temperature that can be relevant, e.g., for neutron stars. In such situation, both $n$ and $K$ are universal physical constants \cite{1964T-GRP}. The polytropic law assumption enables to describe basic properties of the fluid configurations governed by the relativistic laws. The equation of state of the ultrarelativistic degenerate Fermi gas is determined by the polytropic equation with the adiabatic index $\Gamma = 4/3$ corresponding to the polytropic index $n = 3$, while the non-relativistic degenerate Fermi gas is determined by the polytropic equation of state with $\Gamma = 5/3$, and $n = 3/2$ \cite{1983ST-BHW}.

The structure equations of the general relativistic polytropic spheres are determined by the Einstein field equations
\begin{equation}
    R_{\mu\nu} - \frac{1}{2}Rg_{\mu\nu} = \frac{8\pi G}{c^4} T_{\mu\nu},
\end{equation}
and by the local energy-momentum conservation law
\begin{equation}
    T^{\mu\nu}_{\hphantom{\mu\nu};\nu} = 0.
\end{equation}

The structure of the polytropic spheres is governed by the two structure functions. The first one, $\theta(r)$, is related to the mass-energy density radial profile $\rho(r)$ and the central density $\rho_{\mathrm{c}}$ \cite{1964T-GRP}
\begin{equation}
    \rho = \rho_{\mathrm{c}} \theta^{n},
\end{equation}
with the boundary condition $\theta(r=0)=1$. The second one is the mass function given by the relation
\begin{equation}
    m(r) =  \int^{r}_{0} {4 \pi r^{2} \rho \mathrm{d}r},
\end{equation}
with the integration constant chosen to be $m(0) = 0$, to guarantee the smooth spacetime geometry at the origin \cite{1973MTW-G}. At the surface of the configuration at $r=R$, there is $\rho(R)=p(R)=0$, the total mass of the polytropic configuration $M=m(R)$. Outside the polytropic configuration, the spacetime is described by the vacuum Schwarzschild metric.

The structure equations of the polytropic spheres related to the two structure functions, $\theta(r)$ and $m(r)$, and the parameters $n$, $\sigma$, can be put into the form \cite{1964T-GRP,2016SHN-GRP}
\begin{align}
    \frac{\sigma(n+1)}{1+\sigma\theta}\,r\,\frac{\mathrm{d}\theta}{\mathrm{d}r} \left(1-\frac{2Gm(r)}{c^{2}r}\right) + \frac{Gm(r)}{c^{2}r} &= -\frac{G}{c^{2}}\sigma\theta\frac{\mathrm{d}m}{\mathrm{d}r},   \label{grp24}\\
    \frac{\mathrm{d}m}{\mathrm{d}r} &= 4\pi r^{2} \rho_{\mathrm{c}}\theta^{n}.    \label{grp25}
\end{align}

Introducing the characteristic length scale $\mathcal{L}$ of the polytropic sphere \cite{1964T-GRP}
\begin{equation}
    \mathcal{L} = \left[\frac{(n+1)K\rho_{\mathrm{c}}^{1/n}}{4\pi G\rho_{\mathrm{c}}}\right]^{1/2} = \left[\frac{\sigma(n+1)c^{2}}{4\pi G\rho_{\mathrm{c}}}\right]^{1/2},
\end{equation}
and the characteristic mass scale $\mathcal{M}$ of the polytropic sphere
\begin{equation}
    \mathcal{M} = 4\pi \mathcal{L}^3 \rho_\mathrm{c} = \frac{c^2}{G}\sigma (n+1)\mathcal{L},
\end{equation}
the structure equations, Eqs.~(\ref{grp24}) and (\ref{grp25}), can be transformed into dimensionless form by introducing a dimensionless radial coordinate
\begin{equation}
    \xi = \frac{r}{\mathcal{L}},
\end{equation}
and dimensionless gravitational mass function
\begin{align}
    v(\xi) &= \frac{m(r)}{4\pi \mathcal{L}^{3}\rho_{\mathrm{c}}}.
\end{align}

The dimensionless structure equations then take the form (for details see \cite{2016SHN-GRP,1964T-GRP})
\begin{align}
    \xi^{2}\frac{\mathrm{d}\theta}{\mathrm{d}\xi}\frac{1-2\sigma(n+1)\left(v/\xi\right)}{1+\sigma\theta} + v(\xi)&= - \sigma\xi\theta\frac{\mathrm{d}v}{\mathrm{d}\xi}, \label{grp31}\\
    \frac{\mathrm{d}v}{\mathrm{d}\xi} &= \xi^{2}\theta^{n}. \label{grp32}
\end{align}
For fixed parameters $n$, $\sigma$, the structure equations (\ref{grp31}) and (\ref{grp32}) have to be simultaneously solved under the boundary conditions
\begin{equation}
    \theta(0) = 1, \quad v(0) = 0.    \label{grp33}
\end{equation}
From Eqs.~(\ref{grp32}) and (\ref{grp33}) it follows that $v(\xi) \sim \xi^{3}$ for $\xi \to 0$ and, according to Eq.\,(\ref{grp31}), we obtain
\begin{equation}
    \lim_{\xi\to 0_+}\frac{\mathrm{d}\theta}{\mathrm{d}\xi} = 0.
\end{equation}
The surface of the polytropic sphere, $r = R$, is represented by the first zero point of $\theta(\xi)$, denoted as $\xi_{1}$:
\begin{equation}
    \theta(\xi_{1}) = 0.
\end{equation}
Therefore, the solution $\xi_{1}$ determines the surface radius of the polytropic sphere, and the solution $v(\xi_{1})$ determines its gravitational mass.

The solutions of the polytropic structure equations can be obtained by numerical methods only \cite{1964T-GRP}, with the exception of the $n=0$ polytropes governing the spheres with a uniform distribution of the energy density when the solution can be given in terms of the elementary functions \cite{2016SHN-GRP,2000S-SSS}.


\section{Characteristics of the polytropic spheres}
A polytropic sphere constructed for given parameters $n$, $\sigma$ and $\rho_\mathrm{c}$ is characterized by two solutions of the structure equations $\xi_{1}$ and $v(\xi_{1})$ and by the scale factors $\mathcal{L}$ and $\mathcal{M}$. Then the radius of the polytropic sphere reads
\begin{equation}
    R = \mathcal{L} \xi_{1},
\end{equation}
while the gravitational mass of the sphere is given by
\begin{equation}
    M = \mathcal{M} v(\xi_1) = \frac{c^{2}}{G} \mathcal{L}\sigma(n+1) v(\xi_{1}).
\end{equation}

The radial profiles of the energy density, pressure, and mass-distribution are given by the relations
\begin{eqnarray}
  \rho(\xi) &=& \rho_{\mathrm{c}}\theta^{n}(\xi),\\
  p(\xi) &=& \sigma\rho_{\mathrm{c}}\theta^{n+1}(\xi),\\
  M(\xi) &=& M\frac{v(\xi)}{v(\xi_{1})}.
\end{eqnarray}
The temporal metric coefficient takes the form
\begin{equation}
    \mathrm{e}^{2\Phi_\mathrm{int}} = (1+\sigma\theta)^{-2(n+1)} \left\{1-2\sigma(n+1) \frac{v(\xi_{1})}{\xi_{1}} \right\},
\end{equation}
and the radial metric coefficient takes the form
\begin{equation}
    \mathrm{e}^{-2\Psi_\mathrm{int}} = 1 - 2\sigma(n+1) \frac{v(\xi)}{\xi}.
\end{equation}
Detailed discussion of the polytropic spheres, including their gravitational binding energy and the internal energy, can be found in \cite{2016SHN-GRP,1964T-GRP}.

The compactness parameter governing effectiveness of the gravitational binding of the polytropic spheres is given by the relation
\begin{equation}
    \mathcal{C} \equiv \frac{GM}{c^{2}R} = \frac{1}{2}\frac{r_{\mathrm{g}}}{R} = \frac{\sigma(n+1)v(\xi_{1})}{\xi_{1}},
\end{equation}
where we have introduced the standard gravitational radius of the polytropic sphere that reflects its gravitational mass in length units,
\begin{equation}
    r_{\mathrm{g}} = \frac{2GM}{c^{2}}.
\end{equation}
The compactness $\mathcal{C}$ of the polytropic sphere can be represented by the gravitational redshift of radiation emitted from the surface of the polytropic sphere \cite{2011HS-PNR}.

All the characteristic functions introduced above can be determined only by numerical procedures for the polytropic equations of state with $n>0$. The special case of polytropes with $n=0$ corresponds to the physically unrealistic polytropic configurations with a uniform distribution of energy density; for them the characteristic functions can be given in terms of elementary functions and they could serve as a test bed for more complex general polytropes \cite{2000S-SSS,2016SHN-GRP}.

The external vacuum of the polytropic sphere is represented by the Schwarzschild spacetime with the same gravitational mass parameter $M$ as those characterizing the internal spacetime of the polytropic sphere, and is given by the metric coefficients
\begin{equation}
    \mathrm{e}^{2\Phi_\mathrm{ext}} = \mathrm{e}^{-2\Psi_\mathrm{ext}} = 1 - \frac{2GM}{c^{2} r}.
\end{equation}

The photon sphere of the Schwarzschild spacetime, given by the photon circular geodesics, is located at the radius \cite{1973MTW-G}
\begin{equation}
   r_\mathrm{ph} = \frac{3GM}{c^{2}} = \frac{3}{2}r_\mathrm{g}.
\end{equation}
In the following we compare the radius of obtained polytropic spheres containing a region of trapped null geodesics to this radius of photon sphere in order to test, if the original definition of the extremely compact objects ($R<r_\mathrm{ph}$) is satisfied. The relevant condition then reads $C > 1/3$.

In order to have a deeper insight into the character of polytropes containing a region of trapped null geodesics, we will consider also the locally defined compactness of the polytrope, related to a given radius $r = \mathrm{L}\xi$ and given by the relation
\begin{equation}
    \mathcal{C(\xi)} \equiv \frac{\sigma(n+1)v(\xi)}{\xi}.
\end{equation}
We can then test, if the condition $C(\xi) > 1/3$ is satisfied inside the trapping polytropes.


\section{Embeddings of the optical geometry related to the polytropic spheres}
We concentrate our attention on the visualization of the structure of the internal spacetime of the general relativistic polytropes, considering the optical geometry of the spacetime. Such a visualization enables us to find easily the polytropic structures containing a region of trapped null geodesics.

\subsection{Embedding diagrams}
The curvature of the internal spacetime of the polytropes can conveniently be represented by the standard embedding of 2D, appropriately chosen, spacelike surfaces of the ordinary 3-space of the geometry (here, these are $t=\mathrm{const}$ sections of the central planes) into 3D Euclidean space~\cite{1973MTW-G}.

The 3D optical reference geometry~\cite{1988ACL-ORG} related to the spacetime under consideration, enables us to introduce a natural ``Newtonian'' concept of gravitational and inertial forces, reflecting some hidden properties of the test particle motion~\cite{1990A-CFF,1993AMS-CRG,2000SHJ-ORG,2007KS-ORG}. For an alternative approach to the concept of inertial forces see, e.g., the ``special relativistic'' one~\cite{1995S-WFD}. Properties of the inertial forces are reflected by the embedding diagrams of appropriate 2D sections of the optical geometry.  The embedding diagrams of the $n=0$ polytropes were presented in~\cite{2001SHS-NGE}, here they are applied for relativistic polytropes with $n>0$. Note that using the optical reference geometry, it can be shown that extremely compact configurations allowing the existence of bound null geodesics exist \cite{2001SHS-NGE,2009STH-NTE}. For the extremely compact relativistic polytropes with trapped null geodesics a turning point of the embedding diagram of the optical geometry occurs~\cite{2000SHJ-ORG}.

We embed the equatorial plane of the optical reference geometry into the 3D Euclidean space with the line element
\begin{equation}
    \mathrm{d}\tilde{\sigma}^{2} = \mathrm{d}\rho^{2} + \rho^{2} \mathrm{d}\alpha^{2} + \mathrm{d}z^{2}.
\end{equation}
The embedding is represented by a rotationally symmetric surface $z=z(\rho)$ with the 2D line element:
\begin{equation}
    \mathrm{d}\ell_{\mathrm{(E)}}^{2} = \left[1+ \left(\frac{\mathrm{d}z}{\mathrm{d}\rho}\right)^{2} \right] \mathrm{d}\rho^{2} + \rho^{2} \mathrm{d}\alpha^{2}.
\end{equation}

\subsection{Optical reference geometry}
In the static spacetimes, the metric coefficients of the optical 3D space are determined by \cite{1988ACL-ORG}
\begin{equation}
    h_{ik}= \frac{g_{ik}}{-g_{tt}}.   \label{e8}
\end{equation}
In the equatorial plane, the line element has the form
\begin{equation}
    \mathrm{d}\ell^{2}_{(\mathrm{opt})} = h_{rr} \mathrm{d}r^{2} + h_{\phi\phi} \mathrm{d}\phi^{2}    \label{e9}
\end{equation}
that has to be identified with $\mathrm{d}\ell^{2}_{\mathrm{(E)}}$. The azimuthal coordinates of the Optical space and the Euclidean space can be identified ($\alpha \equiv \phi$), but the radial coordinates are related by
\begin{equation}
    \rho^{2} = h_{\phi \phi}.
\end{equation}
Then the embedding formula is determined by
\begin{equation}
    \left(\frac{\mathrm{d}z}{\mathrm{d}\rho}\right)^2 = h_{rr} \left(\frac{\mathrm{d}r}{\mathrm{d}\rho}\right)^{2} -1.
\end{equation}
We transform the embedding formula into a parametric form $z(\rho)= z(r(\rho))$ implying
\begin{equation}
    \frac{\mathrm{d}z}{\mathrm{d}r} = \sqrt{h_{rr}- \left(\frac{\mathrm{d}\rho}{\mathrm{d}r}\right)^{2}}.
\end{equation}
The turning points of the embedding diagrams are given by the condition \cite{2000SHJ-ORG}
\begin{equation}
    \frac{\mathrm{d}\rho}{\mathrm{d}r} = 0.
\end{equation}
We have to include into consideration also the so-called reality condition determining the limits of embeddability
\begin{equation}
    h_{rr} - \left(\frac{\mathrm{d}\rho}{\mathrm{d}r}\right)^{2} \geq 0.
\end{equation}

\subsection{Embeddings of the polytrope optical geometry}
For the general relativistic polytropes, the metric coefficients of the optical geometry take the form
\begin{multline}
    h_{rr}  = \frac{\mathrm{e}^{2\Psi}}{\mathrm{e}^{2\Phi}} = \frac{[1+ \sigma \theta (\xi)]^{2(n+1)}}{1-2\sigma(n+1)v(\xi_{1})/\xi_{1}}\\
    \times \left\{1- 2\sigma (n+1)\frac{v(\xi)}{\xi}\right\}^{-1},
\end{multline}
\begin{equation}
    h_{\phi \phi} = \frac{r^{2}}{\mathrm{e}^{2 \Phi}} = \frac{r^{2}[1+ \sigma \theta (\xi)]^{2(n+1)}}{1-2 \sigma(n+1) v(\xi_{1})/\xi_{1}}.
\end{equation}
It is convenient to introduce a new dimensionless coordinates $\eta$ and $\tilde{z}$ by
\begin{equation}
    \eta = \frac{\rho}{\mathcal{L}},\qquad \tilde{z} = \frac{z}{\mathcal{L}}
\end{equation}
Then we can write
\begin{equation}
    \eta = \frac{\xi[1+ \sigma \theta(\xi)]^{n+1}}{\left[1- 2 \sigma(n+1) \frac{v(\xi_{1})}{\xi_{1}}\right]^{1/2}}
\end{equation}
and
\begin{equation}
    \frac{\mathrm{d}\eta}{\mathrm{d}\xi} = \frac{[1+ \sigma \theta(\xi)]^n \left\{1+ \sigma \left[\theta(\xi) + (n+1) \xi \frac{\mathrm{d}\theta}{\mathrm{d}\xi}\right]\right\}}{\left[1- 2 \sigma(n+1) \frac{v(\xi_{1})}{\xi_{1}}\right]^{1/2}}.  \label{20}
\end{equation}
The condition governing the turning points of the embedding diagrams reads
\begin{equation}
    \sigma\left[\theta(\xi) + (n+1) \xi \frac{\mathrm{d}\theta}{\mathrm{d}\xi}\right] = -1.   \label{e21}
\end{equation}
The embedding formula takes the form

\begin{widetext}
\begin{multline}
    \left(\frac{\mathrm{d}\tilde{z}}{\mathrm{d}\xi}\right)^{2} = \left\{1- 2 \sigma (n+1)\frac{v(\xi_{1})}{\xi_{1}}\right\}^{-1}\left\{1-2\sigma (n+1)\frac{v(\xi)}{\xi}\right\}^{-1} 2\sigma (n+1)[1+ \sigma \theta(\xi)]^{2n}\\
    \times \left\{\left\{1+ \sigma\left[\theta(\xi) + (n+1)\xi\frac{\mathrm{d}\theta}{\mathrm{d}\xi}\right]\right\}^2 \frac{v(\xi)}{\xi} - \xi \frac{\mathrm{d}\theta}{\mathrm{d}\xi}\left[1+ \sigma\theta(\xi)+\frac{\sigma}{2}(n+1)\xi \frac{\mathrm{d}\theta}{\mathrm{d}\xi}\right]\right\}.
\end{multline}
\end{widetext}

The condition of embeddability giving the limits of applicability of the embedding procedure takes the form
\begin{multline}
    \left\{1+ \sigma \left[\theta(\xi) + (n+1) \xi \frac{\mathrm{d}\theta}{\mathrm{d}\xi}\right] \right\}^2 \frac{v(\xi)}{\xi}\\
     - \xi \frac{\mathrm{d}\theta}{\mathrm{d}\xi} \left[1+ \sigma \theta (\xi) + \frac{\sigma}{2} (n+1) \xi \frac{\mathrm{d}\theta}{\mathrm{d}\xi} \right]\geq 0.
\end{multline}

\begin{figure}
\centering\includegraphics[width=0.7\linewidth,keepaspectratio=true]{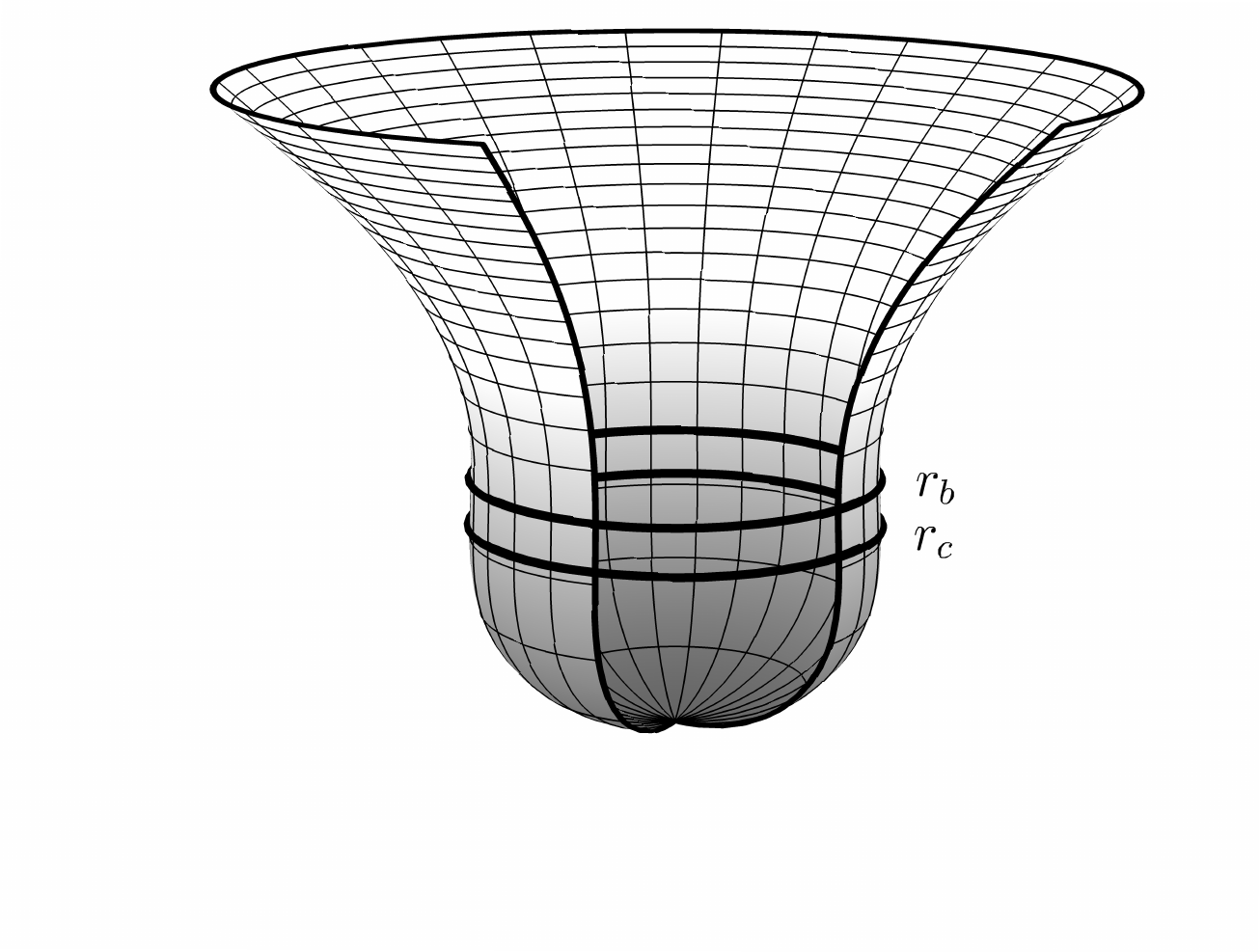}%
\caption{\label{fig1} Optical geometry of the polytropic sphere having $n=3$ and $\sigma = 3/4$. Turning points corresponding to the stable and unstable null circular geodesics are depicted as circles $r_\mathrm{c}$ and $r_\mathrm{b}$.}
\end{figure}

Notice that the embedding diagrams are related purely to the solutions of the dimensionless structure equations of the general relativistic polytropes, being independent of the length scale factor $\mathcal{L}$ governing physical extension and gravitational mass of the polytropes. For this reason, the existence of the zones of null geodesics trapping will be also independent of the length scale factor $\mathcal{L}$. The trapping phenomenon is thus fully governed by the polytrope parameters $n$ and $\sigma$ --- it is formally independent of the central density $\rho_\mathrm{c}$ that, however, enters definition of the relativistic parameter $\sigma$.

\begin{figure}
\centering\includegraphics[width=0.99\linewidth,keepaspectratio=true]{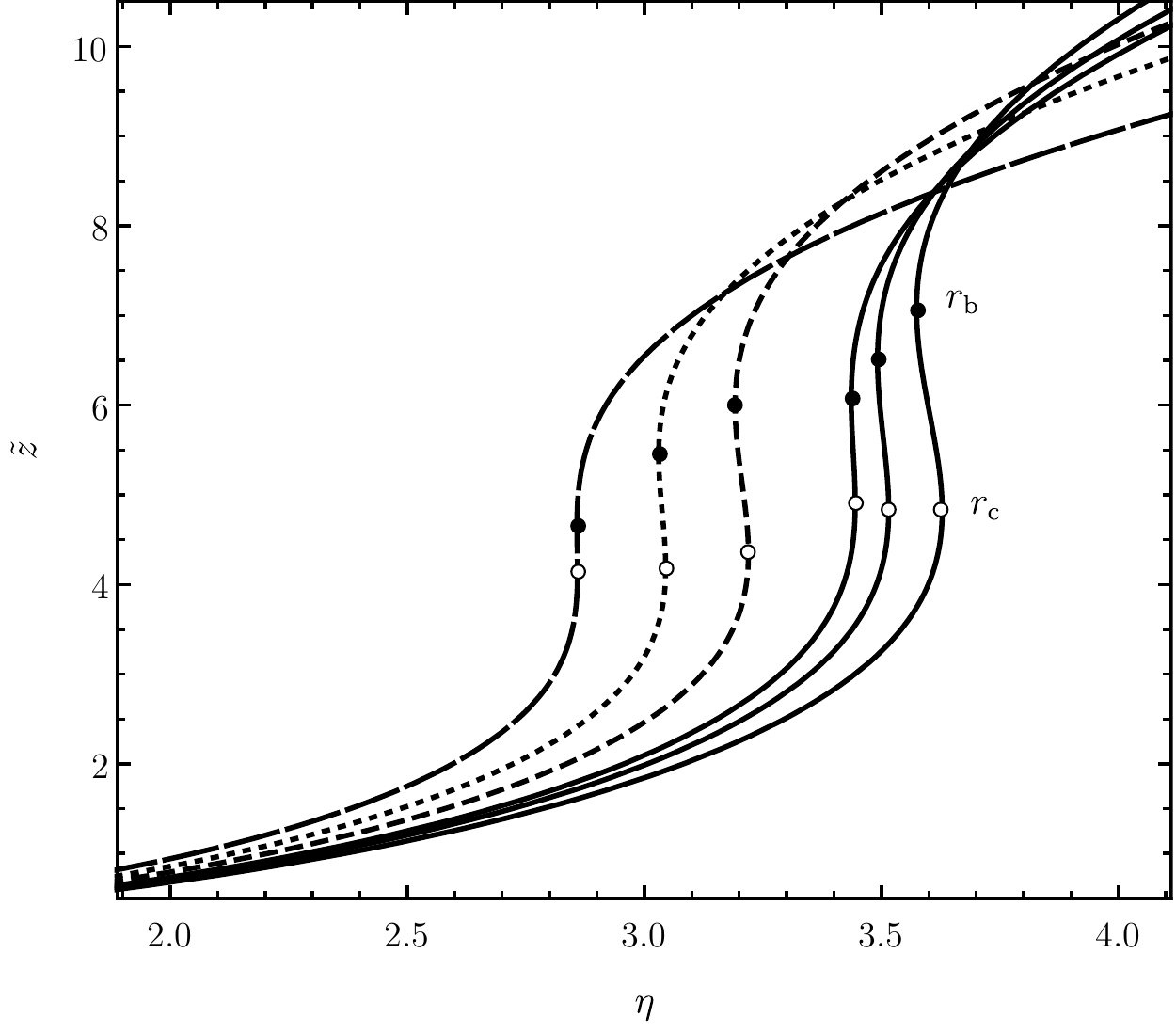}%
\caption{\label{fig2} Embedding diagram constructed for several polytropic spheres having the polytropic and relativistic parameters $\{n,\sigma\}$ valued gradually as $\{2.2, 11/16\}$, $\{2.5,5/7\}$, $\{2.7,27/37\}$, $\{3,7/10\}$, $\{3,18/25\}$, $\{3,3/4\}$ (curves in same order as growing $\eta$ coordinate of the turning points $r_\mathrm{c}$ and $r_\mathrm{b}$ corresponding to null circular geodesics).}
\end{figure}

Using numerically obtained solutions of the polytrope structure equations, we give examples of the embedding diagrams. In Fig.~\ref{fig1} the embedding diagram of the optical geometry of the internal $n = 3$ polytrope spacetime is given for the extremal value of the relativistic parameter $\sigma = 3/4$ allowed by the causality limit \cite{1964T-GRP}. As demonstrated in \cite{2000SHJ-ORG}, the turning points of the diagram correspond to the (inner) stable null circular geodesics at radius $r_\mathrm{ph(s)} = r_\mathrm{c}$ and the (outer) unstable null circular geodesic at radius $r_\mathrm{ph(u)} = r_\mathrm{b}$. The radius $r_\mathrm{c}$ corresponds to the centre of the trapping region, while the radius $r_\mathrm{b}$ corresponds to its outer boundary. In Fig.~\ref{fig2}, we demonstrate how the optical geometry embeddings depend on the polytrope index $n$, and on the relativistic parameter $\sigma$ for fixed $n$. In the next section, we use the embeddings for detailed study of the existence of the trapping zones for null geodesics in dependence on the polytropic parameters $n$ and $\sigma$.

\section{General relativistic polytropes containing region of trapped null geodesics}
Numerical solutions of the structure equations of the polytrope spheres yield the dimensionless radial profiles of energy density, mass and metric coefficients, and the dimensionless extension and mass parameters $\xi_1$ and $v_1 = v_{\xi_1}$. These solutions are governed by the parameters $n$ and $\sigma$, being independent of the third parameter governing the polytrope spheres, $\rho_\mathrm{c}$, that governs the length and mass scales of the polytropes. We restrict our attention to the polytropic spheres with the standard choice of the polytropic index, $0 \leq n \leq 4$.

\begin{figure}
\centering\includegraphics[width=0.99\linewidth,keepaspectratio=true]{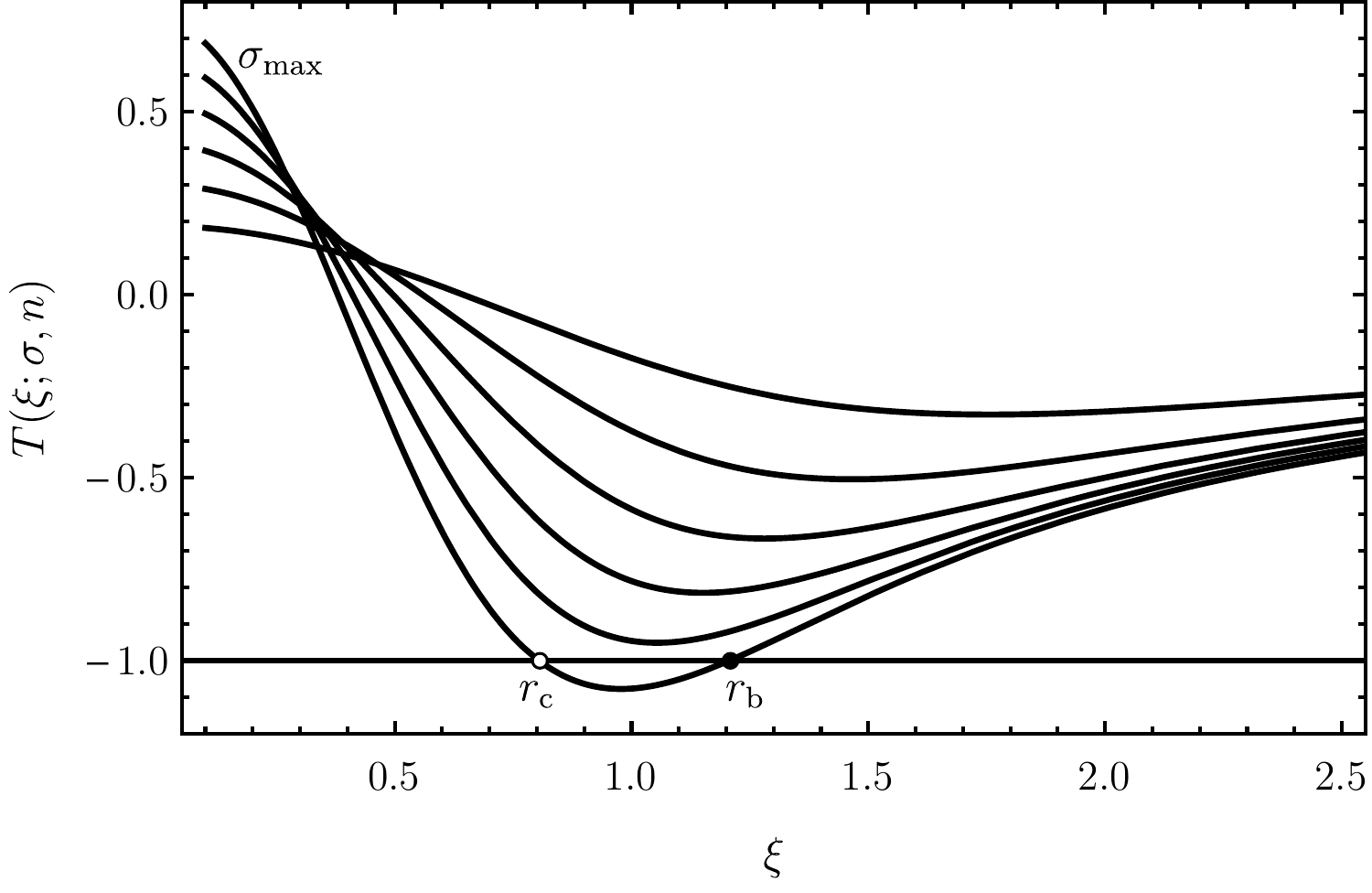}%
\caption{\label{fig3} Searching of the critical relativistic parameter $\sigma_{\mathrm{min}}(n)$ and the turning points of the ``turning'' optical geometry embedding diagrams. If turning function $T(\xi; \sigma, n) = -1$ for two different values of coordinate $\xi$, the trapping effect exists for given pair $\{n, \sigma\}$. Curves are depicted for $n=3$. Parameter $\sigma$ is gradually increased from given nonzero value up to $\sigma_\mathrm{max}$. Similar behaviour of plotted turning functions can be seen also for other values of the polytropic index  $n>2.138$. }
\end{figure}

\subsection{Demarcation of trapping region in the $n$--$\sigma$ parameter space}
As demonstrated in the previous section, a region with trapped null geodesics can exists in the interior of the polytropic spheres, if the parameters $n$ and $\sigma$ are conveniently chosen. We thus give first the region of the $n$--$\sigma$ parameter space determining the polytropic spheres demonstrating the trapping phenomenon.

We have to put in the beginning the upper causal limit on the relativistic parameter. To avoid a super-luminal speed of sound in the gas, maximal value of the relativistic parameter $\sigma$ for fixed polytropic index $n$ is limited. For adiabatic processes in the polytropic spheres, the phase velocity of the sound is given by
\begin{equation}
    v_\mathrm{s}^2=\left(\frac{\mathrm{d}p}{\mathrm{d}\varrho}\right)_\mathrm{adiabatic}.
\end{equation}
Because the radial profile of the pressure in any polytropic fluid sphere is a monotonically decreasing function, the limit on maximum value of $\sigma$ results from the restriction on the speed of sound in the centre, giving thus the relation
\begin{equation}
    v_\mathrm{sc}\equiv c\left(\frac{n+1}{n}\sigma\right)^{1/2} < c.
\end{equation}
Whence for a given polytropic index $n$ one gets the upper limit restriction
\begin{equation}
    \sigma \leq \frac{n}{n+1}\equiv \sigma_\mathrm{max}.
\end{equation}

\begin{figure}
\centering\includegraphics[width=0.99\linewidth,keepaspectratio=true]{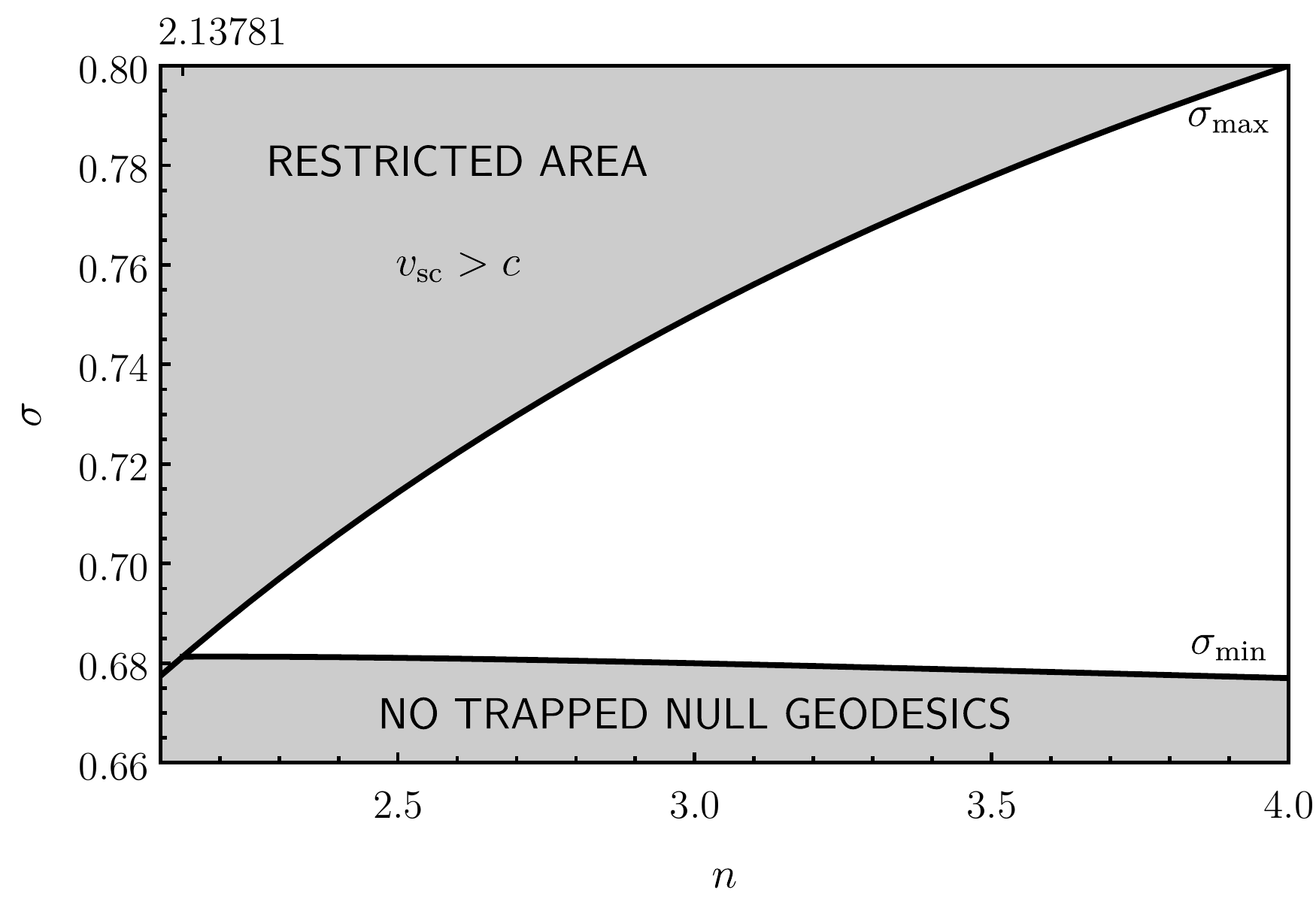}%
\caption{Trapping polytropes in parameter $n$-$\sigma$ space. Figure is giving the span of possible parameter $\sigma$ for given polytropic index in the interval $2.138 < n < 4$ for which the trapping effect exist. \label{fig4}}
\end{figure}

Using the behaviour of left hand side of Eq.~(\ref{e21}), denoted as the turning function $\sigma\left[\theta(\xi) + (n+1) \xi \frac{\mathrm{d}\theta}{\mathrm{d}\xi}\right] \equiv T(\xi;\sigma,n)$, we can numerically search for the existence of polytropic spheres demonstrating the trapping effect by solving the equation $T(\xi;\sigma,n) = -1$, as graphically depicted in Fig.~\ref{fig3} in the special case of $n=3$ polytropes. If there are, for a fixed $n$, some values of $\sigma$ implying two different solutions, $\xi_\mathrm{c}$ and $\xi_\mathrm{b}$, of this equation, the trapping region exists, and the solutions give radii of the stable and unstable circular geodesics. We shall confirm this conclusion in the following by direct study of the effective potential of the null geodesics of the internal spacetime of such polytropic configurations. If there is, for the fixed $n$, only one solution, where $\xi_\mathrm{c} = \xi_\mathrm{b}$, the minimal value of the relativistic parameter $\sigma_\mathrm{min}$ allowing for trapping is found. The numerical analysis demonstrates that the trapping region start to exist for properly selected relativistic parameter $\sigma$, if the polytropic index overcomes the critical minimal value of $n_\mathrm{min} \doteq 2.1378$. The limiting maximal (and simultaneously minimal) allowed value of the relativistic parameter reads $\sigma_\mathrm{max}(n=2.1378) = 0.681$.

\begin{figure}
\centering\includegraphics[width=0.99\linewidth,keepaspectratio=true]{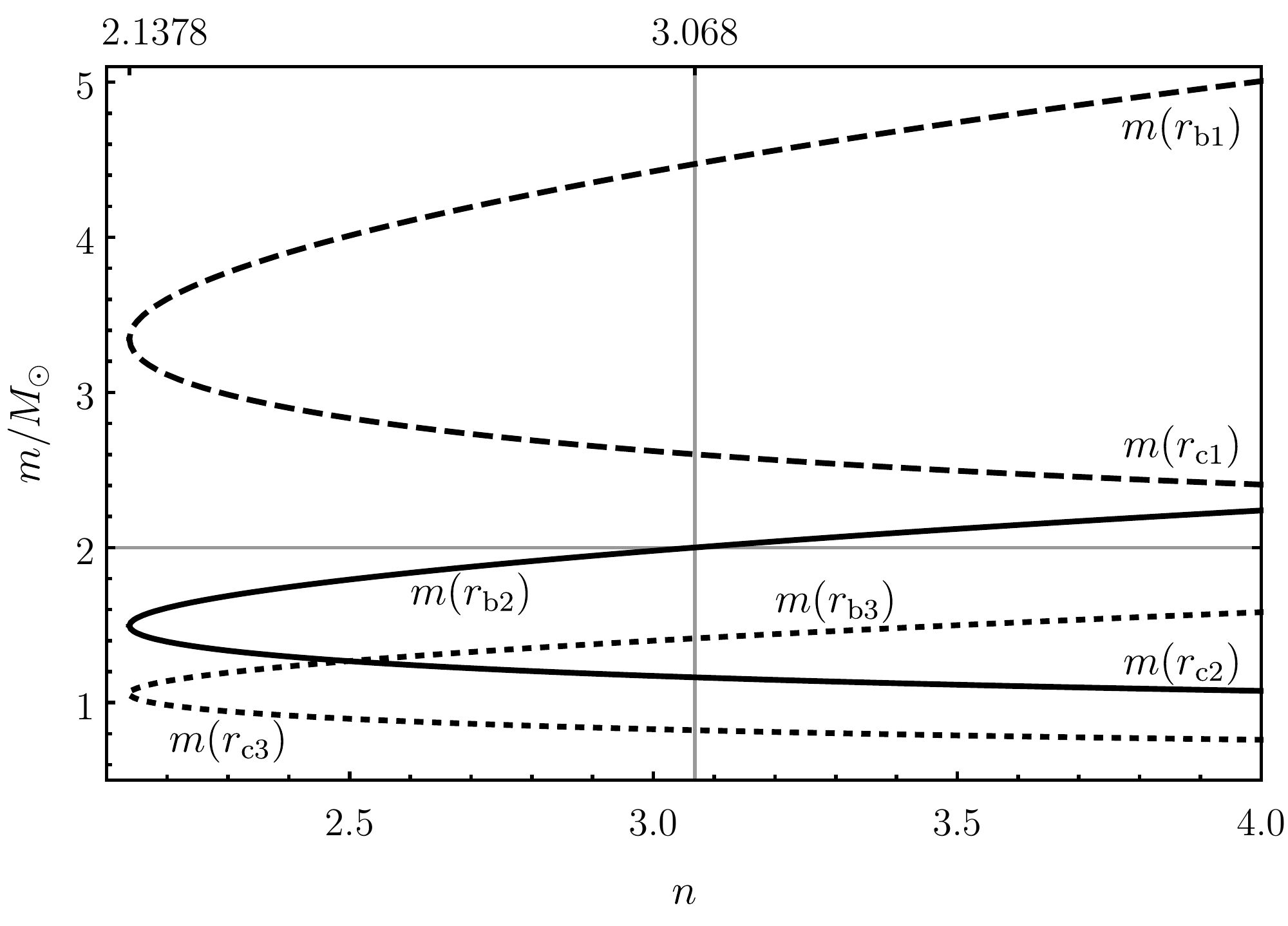}%
\caption{Mass of the trapping polytropes with $\sigma = \sigma_\mathrm{max}$, contained under the radius of photon circular orbits $r_\mathrm{c}$ and $r_\mathrm{b}$, is given in dependence on the polytropic index $n$ for three central densities $\varrho_\mathrm{c}$: 1) $10^{15}$~g\,cm$^{-3}$ (dashed lines); 2) $5\times 10^{15}$~g\,cm$^{-3}$ (solid lines); 3) $10^{16}$~g\,cm$^{-3}$ (dotted lines).   \label{fig5}}
\end{figure}

\begin{figure}
\centering\includegraphics[width=0.99\linewidth,keepaspectratio=true]{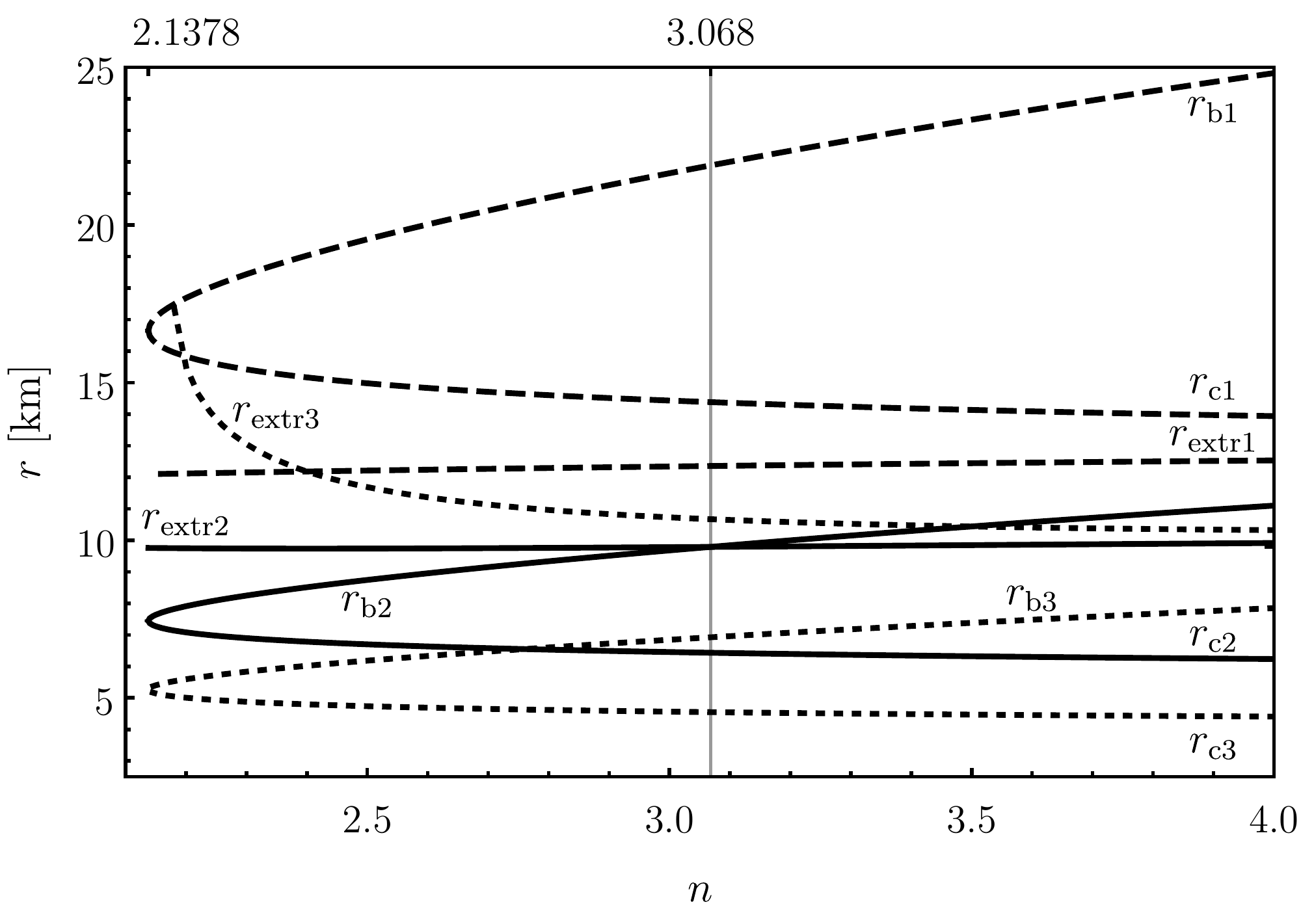}%
\caption{Position of the null geodesics stable $r_\mathrm{c}$ and unstable $r_\mathrm{b}$ radii and the limit radius $r_\mathrm{extr}$ in dependence on polytropic index $n$ is given for the polytropes with three central densities $\varrho_\mathrm{c}$: 1) $10^{15}$~g\,cm$^{-3}$ (dashed lines); 2) $5\times 10^{15}$~g\,cm$^{-3}$ (solid lines); 3) $10^{16}$~g\,cm$^{-3}$ (dotted lines).  For the last case and $n < 2.17$, the polytropic balls have mass $M < 2M_\odot$, therefore $r_\mathrm{extr3}$ does not exist for such values of $n$. \label{fig6}}
\end{figure}

\begin{figure}
\centering\includegraphics[width=0.99\linewidth,keepaspectratio=true]{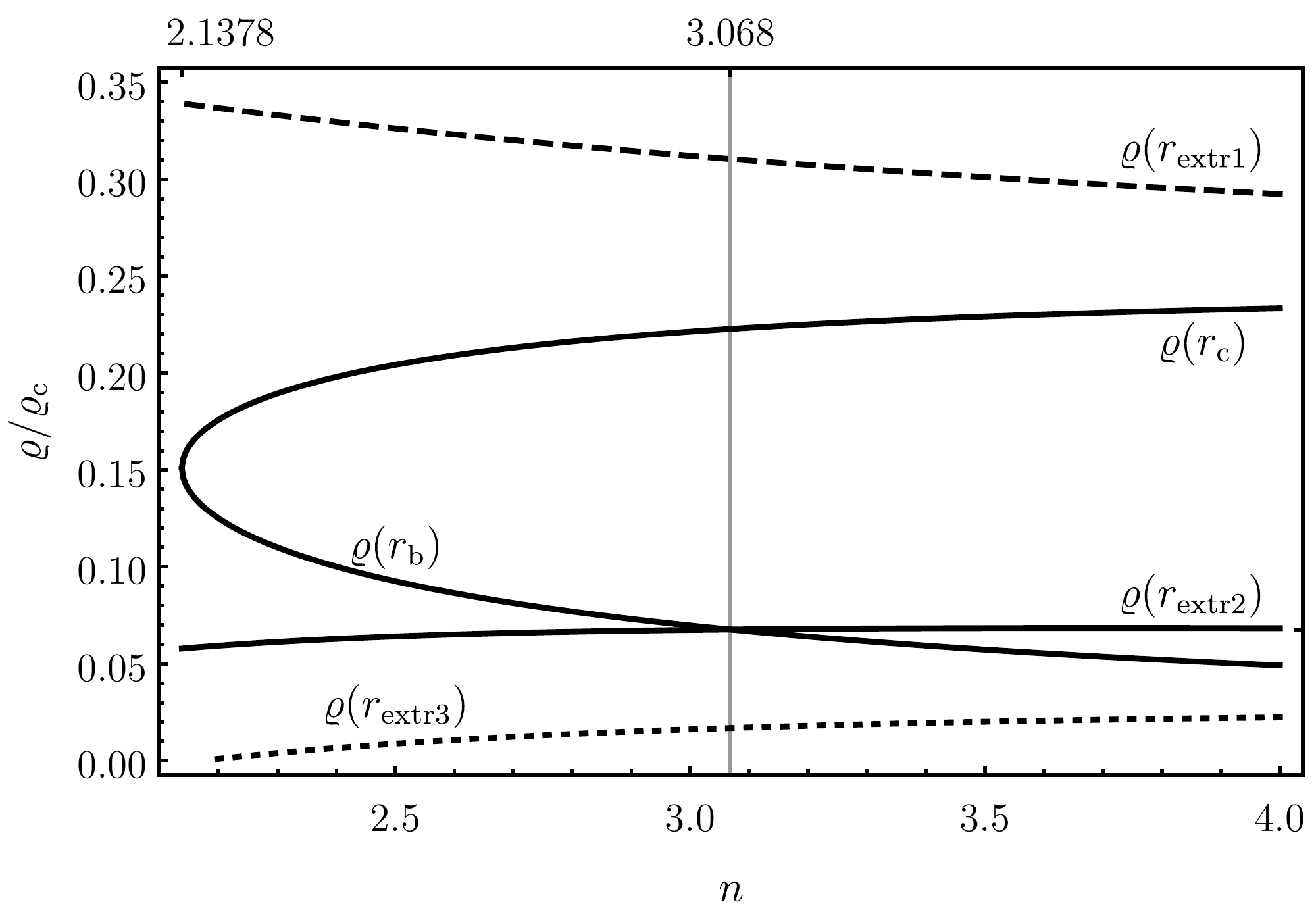}%
\caption{Trapping polytropes with $\sigma = \sigma_\mathrm{max}$. Energy density on location of null geodesics and loci of $r_\mathrm{extr}$ in dependence on the polytropic index $n$ are given for three central densities $\varrho_\mathrm{c}$: 1) $10^{15}$~g\,cm$^{-3}$ (dashed lines); 2)~$5\times 10^{15}$~g\,cm$^{-3}$ (solid lines); 3) $10^{16}$~g\,cm$^{-3}$ (dotted lines). Functions $\varrho (r_\mathrm{c})/\varrho_\mathrm{c}$ and $\varrho (r_\mathrm{c})/\varrho_\mathrm{c}$ are the same for all three depicted cases. For the last case and $n < 2.17$ the polytropic balls have mass $M < 2M_\odot$.   \label{fig7}}
\end{figure}

Based on the above given procedure, the turning point limit $\sigma_\mathrm{min}(n)$ presented in Fig.~\ref{fig4} has been obtained. Together with the causal limit of $\sigma_\mathrm{max}(n)$, the turning points limit provides in the parameter $n$--$\sigma$ plane restriction on the existence of polytrophic spheres containing trapped null geodesics.

\subsection{Physically relevant polytropic spheres}
Now we can provide more detailed information on the physical properties of the polytropic spherical configurations containing regions of trapped null geodesics. For selected values of the polytropic index $n$ and related maximal allowed values of the relativistic parameter $\sigma_\mathrm{max}(n)$, properties of such configurations, like the total gravitational mass, the surface radius, and the radii $r_\mathrm{c}$ and $r_\mathrm{b}$ governing the trapping zone of the polytrope, are summarized in Tab.~\ref{tab1} for the special selection of the central energy density of the polytropic sphere $\rho_\mathrm{c} = 5 \times 10^{15}$~g\,cm$^{-3}$. We give also the ratio $\rho(r)/\rho_\mathrm{c}$ at the radii $r = r_\mathrm{c}$, $r = r_\mathrm{b}$. Compactness of these polytropes will be studied separately.

\begin{table*}
\caption{List of parameters describing polytropic fluid sphere of given $n$ having $\sigma=\sigma_\mathrm{max}$. For calculation of scale parameter $\mathcal{L}$, value $5 \times 10^{15}$~g\,cm$^{-3}$ was used as $\varrho_\mathrm{c}$. \label{tab1}}
\begin{ruledtabular}
\begin{tabular}{ccccccccccccccc}
 $n$ & $\sigma_\mathrm{max}$ & $R$ [km] & $M/M_\odot$ & $r_\mathrm{c}$ [km]& $\varrho(r_\mathrm{c})/\varrho_\mathrm{c}$ & $r_\mathrm{b}$ [km] & $\varrho(r_\mathrm{b})/\varrho_\mathrm{c}$ & $r_\mathrm{extr}$ [km]&$\varrho(r_\mathrm{extr})/\varrho_\mathrm{c}$\\ \hline

 4.0 & 0.80000 & $1.87\times 10^8$ & 1251.82 & 6.2323 & 0.23344 & 11.0981 & 0.04914& 9.9120 & 0.06835 \\
 3.9 & 0.79592 & $3.99\times 10^7$ & 1102.78 & 6.2476 & 0.23275 & 10.9720 & 0.05055 & 9.8988 & 0.06840 \\
 3.8 & 0.79167 & $9.27\times 10^7$ & 1602.90 & 6.2639 & 0.23200 & 10.8431 & 0.05205 & 9.8854 & 0.06843 \\
 3.7 & 0.78723 & 64276.3 & 54.1217 & 6.2813 & 0.23115 & 10.7111& 0.05367 & 9.8719 & 0.06845 \\
 3.6 & 0.78261 & 21288.2 & 42.8887 & 6.3001 & 0.23021 & 10.5758& 0.05541 & 9.8584 & 0.06843 \\
 3.5 & 0.77778 & 11280.6 & 34.6763 & 6.3203 & 0.22916 & 10.4370 & 0.05730 & 9.8448 & 0.06839 \\
 3.4 & 0.77273 & 7173.57 & 28.0331 & 6.3422 & 0.22798 & 10.2943& 0.05935 & 9.8313 & 0.06831 \\
 3.3 & 0.76744 & 4998.89 & 22.3502 & 6.3661 & 0.22664 & 10.1475& 0.06158 & 9.8179 & 0.06819 \\
 3.2 & 0.76190 & 3598.36 & 17.3233 & 6.3923 & 0.22512 & 9.9961 & 0.06404 & 9.8047 & 0.06801 \\
 3.1 & 0.75610 & 2476.05 & 12.8706 & 6.4211 & 0.22338 & 9.8396 & 0.06674 & 9.7919 & 0.06777 \\
 3.0 & 0.75000 & 1446.86 & 9.21101 & 6.4531 & 0.22138 & 9.6774 & 0.06975 & 9.7796 & 0.06746 \\
 2.9 & 0.74359 & 692.632 & 6.71466 & 6.4890 & 0.21906 & 9.5088 & 0.07313 & 9.7680 & 0.06706 \\
 2.8 & 0.73684 & 325.643 & 5.27798 & 6.5297 & 0.21633 & 9.3327 & 0.07695 & 9.7574 & 0.06655 \\
 2.7 & 0.72973 & 173.764 & 4.45141 & 6.5766 & 0.21308 & 9.1475 & 0.08134 & 9.7481 & 0.06591 \\
 2.6 & 0.72222 & 106.141 & 3.92699 & 6.6317 & 0.20915 & 8.9513 & 0.08646 & 9.7406 & 0.06510 \\
 2.5 & 0.71429 & 71.9182 & 3.55954 & 6.6981 & 0.20429 & 8.7406 & 0.09257 & 9.7356 & 0.06410 \\
 2.4 & 0.70588 & 52.5214 & 3.28159 & 6.7817 & 0.19804 & 8.5094 & 0.10011 & 9.7338 & 0.06285 \\
 2.3 & 0.69697 & 40.5057 & 3.05917 & 6.8949 & 0.18950 & 8.2453 & 0.11002 & 9.7365 & 0.06128 \\
 2.2 & 0.68750 & 32.5327 & 2.87357 & 7.0776 & 0.17591 & 7.9080 & 0.12503 & 9.7454 & 0.05931 \\
\end{tabular}
\end{ruledtabular}
\end{table*}

In the second step of our considerations, we put our results into an astrophysical context, related to the observational restrictions of the neutron stars, by considering dependence of our general dimensionless results on the central energy density parameter $\rho_\mathrm{c}$ governing the extension and mass of the polytrope configuration. It should be noted that all the characteristic radii and masses of the polytropic spheres depend significantly on the central energy density --- with increasing central density the radii and masses decrease as $1/\sqrt{\rho_\mathrm{c}}$.

Because of the observationally given limit on the mass of the neutron stars $M_\mathrm{{max(o)}} \sim 2M_{\odot}$ \cite{2006LP-ESN,2014L-NS}, we will assume that the physical relevance of the polytropic equation of state is limited by the radius $r_\mathrm{extr}$ where the polytrope sphere reaches the mass $m(r_\mathrm{extr}) = 2M_{\odot}$; above this radius the sphere should be described by different equations of state, as well known from the theory of neutron stars \cite{1983ST-BHW}. The maximum mass of neutron stars allowed theoretically by realistic equations of state $M_\mathrm{max(t)} \sim 2.8M_{\odot}$ \cite{2013CHZ-MMN}, while the minimal surface radius $R_\mathrm{min} \sim 10$~km \cite{2016LP-EOS}. Moreover, we will assume that the central energy density is supernuclear, i.e., $\rho_\mathrm{c} > 10^{15}$~g\,cm$^{-3}$.

In order to search for trapping polytropes having physical relevance, we have to compare the characteristic radii of the trapping zone $r_\mathrm{c}$ and $r_\mathrm{b}$ to the $r_\mathrm{extr}$ radius giving limit on physical relevance of the polytropic sphere. This can be done using the results presented in Tab.~\ref{tab1} for the central density $\rho_\mathrm{c} = 5 \times 10^{15}$~g\,cm$^{-3}$. We can see that the gravitational mass and the surface radius of the relativistic polytropes exceed significantly the mass and radii related to the observed neutron stars; the discrepancy strongly increases with increasing polytropic index $n$. On the other hand, the limiting radius $r_\mathrm{extr}$ is for all considered values of $n$ well comparable to the observed radii of neutron stars ($r_\mathrm{extr} < R_\mathrm{min}$). For all values of $n$, the central radius of the trapping zone $r_\mathrm{c} < r_\mathrm{extr}$, guaranteeing thus that the trapping is possible for all considered polytropes. However, there is $r_\mathrm{b} < r_\mathrm{extr}$ only for the polytropes with $n < 3.1$. Only for these polytropic spheres the whole trapping zone will be contained in the allowed region of the polytropic sphere.

We can also see that the energy density at the loci of the stable circular geodesic are on the level of $10^{-1}\rho_\mathrm{c}$, but only on the level of  $10^{-2}\rho_\mathrm{c}$ at the unstable circular null geodesic giving the outer edge of the trapping zone. Of course, the energy ratio depends significantly on the values of the spacetime parameters $n$, $\sigma$, $\rho_\mathrm{c}$, as demonstrated in Tab.~\ref{tab1}. On the other hand the ration $\varrho(r_\mathrm{extr})/\varrho_\mathrm{c} \sim 0.07$ with only slight dependence on $n$.

In order to illustrate the situation of the physical relevance of the trapping zones, we give in Fig.~\ref{fig5} the functions $m(r_\mathrm{c}; n, \sigma_\mathrm{max})$ and $m(r_\mathrm{b}; n, \sigma_\mathrm{max})$, and in Fig.~\ref{fig6}  the functions  $r_\mathrm{c}(n, \sigma_\mathrm{max})$,  $r_\mathrm{b}(n, \sigma_\mathrm{max})$ and  $r_\mathrm{extr}(n, \sigma_\mathrm{max})$. In order to have a good insight into the character of the trapping polytropes, we give also the functions $\rho(r_\mathrm{c})/\rho_\mathrm{c}(n, \sigma_\mathrm{max})$, $\rho(r_\mathrm{b})/\rho_\mathrm{c}(n, \sigma_\mathrm{max})$, and $\rho(r_\mathrm{extr})/\rho_\mathrm{c}(n,\sigma_\mathrm{max})$ in Fig.~\ref{fig7}. We give these functions for three characteristic selections of the central density: $\varrho_\mathrm{c} = 10^{15},\ 5\times 10^{15},\ 10^{16}$~g\,cm$^{-3}$. The numerical results show that whole the trapping zones are located under $r_\mathrm{extr}$ for all the polytropes with $n \leq 4$, if $\rho_\mathrm{c} = 10^{16}$~g\,cm$^{-3}$. On the other hand, whole the trapping zones are located above $r_\mathrm{extr}$ for all the polytropes with $n \leq 4$, if $\rho_\mathrm{c} = 10^{15}$~g\,cm$^{-3}$ --- in this case the trapping effect should be physically implausible. In the intermediate case of $\varrho_\mathrm{c} \sim 5\times 10^{15}$~g\,cm$^{-3}$, the trapping zones can be fully contained in polytropes with $n\leq3$. For $n>3$, the trapping zones reach the polytrope surface at $r_\mathrm{extr}$.

\subsection{Effective potential of null geodesics}
We test validity of our results, based on the study of the embedding diagrams of the optical geometry of the polytropic spheres, by using the direct study of the effective potential of the null geodesics that gives whole the information on the trapping zones \cite{2012SHU-NTE}.

Four-momentum $p^\mu$ of particles moving along null geodesics satisfies the geodesic equation ($\lambda$ is an affine parameter)
\begin{equation}
  \frac{\mathrm{D}p^\mu}{\mathrm{d}\lambda}=0,
\end{equation}
simultaneously with the normalization condition
\begin{equation}\label{normalcond}
  p^\mu p_\mu = 0.
\end{equation}
Because of the spherical symmetry of the studied internal metric, the motion plane is central, and for a single-particle motion it is reasonable to choose the equatorial plane ($\theta = \pi/2$). Moreover, axial symmetry and time independence of the metric induce existence of two Killing vector fields resulting in conserved energy and axial angular momentum of the particle
\begin{equation}
  E = - p_t,\qquad L = p_\psi.
\end{equation}
Then the radial component of the geodesic motion, derived using (\ref{normalcond}), has to fulfill the relation
\begin{equation}\label{radialmot}
  (p^r)^2 = e^{-2(\Phi+\Psi)}E^2\left(1-e^{2\Phi}\frac{\ell^2}{r^2}\right),
\end{equation}
where $\ell \equiv L/E$ is the impact parameter. For both the internal and external spacetime of the polytropic sphere, the turning points of the radial motion can be thus expressed by an effective potential $V_\mathrm{eff}$ with respect to the impact parameter. The motion is then allowed in regions where the impact dimensionless parameter $\tilde{\ell}$ satisfied the condition
\begin{equation}
  {\tilde{\ell}}^2\equiv \frac{\ell^2}{\mathcal{L}^2} \leq V{}_\mathrm{eff}^\mathrm{int/ext}\equiv \frac{\xi^2}{\exp(2\Phi_{\mathrm{int/ext}})}.
\end{equation}
In the polytrope interior with parameters $(n,\sigma)$, the effective potential $V{}^\mathrm{int}_\mathrm{eff}$ is determined by the metric coefficient $g_\mathrm{tt}(\xi;n,\sigma)$ having its radial profile fully governed by the function $\theta(\xi;n,\sigma)$.

Condition for existence of the local maxima and minima of the internal spacetime effective potential, $\mathrm{d}V{}_\mathrm{eff}^\mathrm{int}/\mathrm{d}\xi = 0$, determines the loci of the null circular geodesics in terms of equation
\begin{equation}
  \theta(\xi) + (n+1) \xi \frac{\mathrm{d}\theta(\xi)}{\mathrm{d}\xi} = -\frac{1}{\sigma}. \label{veff1}
\end{equation}
For given polytrope index $n$, the limiting case corresponding to coalescence of the radii of the circular null geodesics at an inflexion point of the effective potential, determined by the condition $\mathrm{d}^2 V{}_\mathrm{eff}^\mathrm{int}/\mathrm{d}\xi^2 = 0$ implying the relation
\begin{widetext}
\begin{equation}
  1 + \sigma \left\{\sigma \theta(\xi)^2+(n+1)\xi\left[\frac{\mathrm{d}\theta(\xi)}{\mathrm{d}\xi}\left(4+(2n+1)\sigma\xi \frac{\mathrm{d}\theta(\xi)}{\mathrm{d}\xi}\right)+\xi\frac{\mathrm{d}^2 \theta(\xi)}{\mathrm{d}\xi^2}\right]+\theta(\xi)\left[2+(n+1)\sigma \xi\left(4\frac{\mathrm{d}\theta(\xi)}{\mathrm{d}\xi} + \xi \frac{\mathrm{d}^2\theta(\xi)}{\mathrm{d}\xi^2}\right)\right]\right\}=0,\label{veff2}
\end{equation}
\end{widetext}
gives the minimal allowed value of the relativistic parameter $\sigma_\mathrm{min}(n)$. For a given $n > 2.138$, allowing existence of the circular null geodesics, we can find the minimal value of the relativistic parameter $\sigma_\mathrm{min}(n)$ due to simultaneous solving of  the condition for an inflexion point of the effective potential, with the extrema relation given by Eq.~(\ref{veff1}), we obtain a simple relation governing $\sigma_\mathrm{min}(n)$ in the form
\begin{equation}
  (n+2)\frac{\mathrm{d}\theta(\xi)}{\mathrm{d}\xi} + (n+1)\xi \frac{\mathrm{d}^2\theta(\xi)}{\mathrm{d}\xi^2} = 0.
\end{equation}
The two solutions of Eq.~(\ref{veff1}) for given $n$ and $\sigma\in \left(\sigma_\mathrm{min},\sigma_\mathrm{max}\equiv n/(n+1)\right)$ give the stable circular null geodesics located at $r_\mathrm{c}$ where $\mathrm{d}^2 V{}_\mathrm{eff}^\mathrm{int}/\mathrm{d}\xi^2 < 0$ and unstable circular null geodesics at $r_\mathrm{b}$ where $\mathrm{d}^2 V{}_\mathrm{eff}^\mathrm{int}/\mathrm{d}\xi^2 > 0$.

\begin{figure}
\centering\includegraphics[width=0.99\linewidth,keepaspectratio=true]{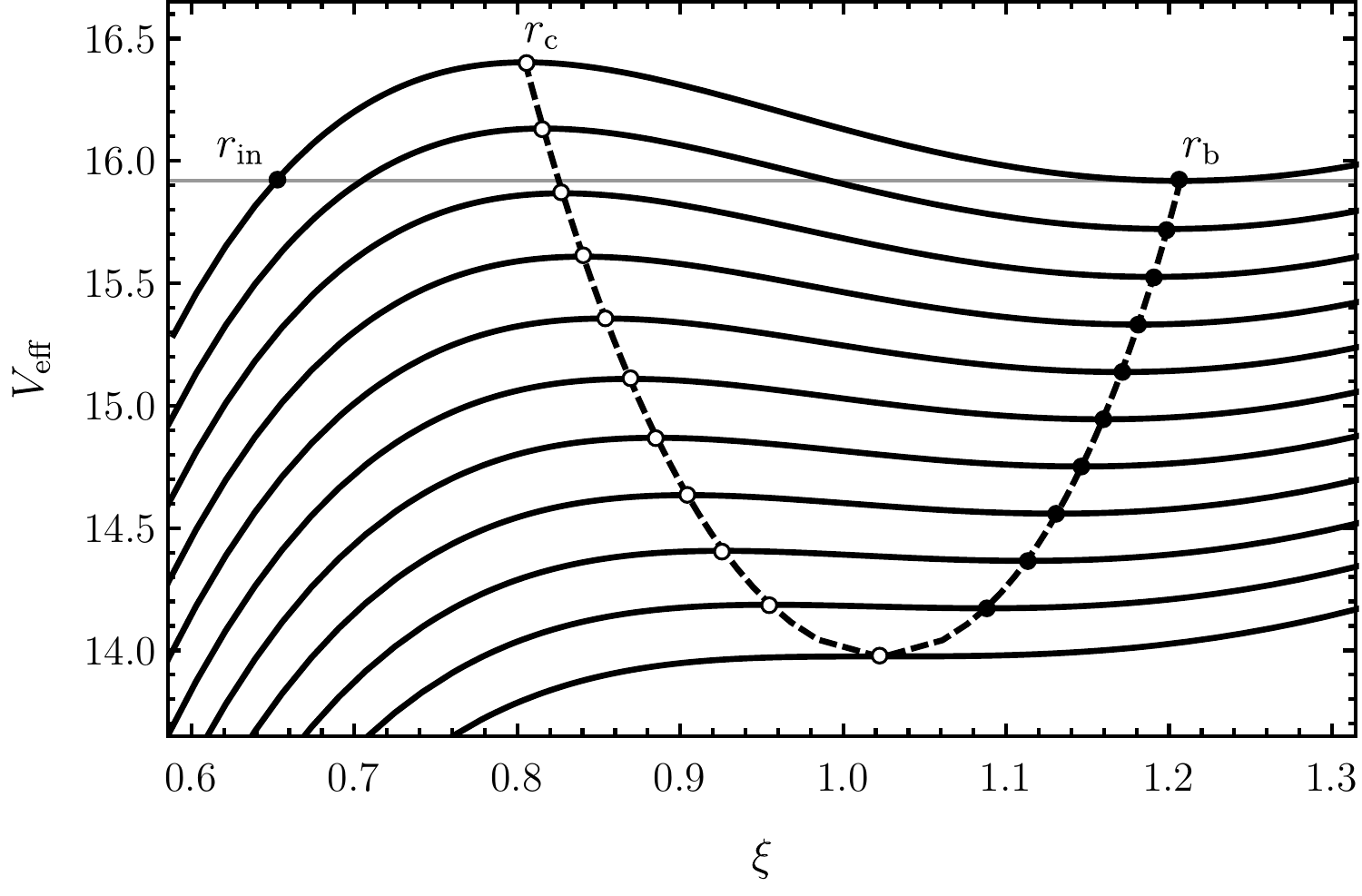}%
\caption{The null geodesic effective potential $V{}_\mathrm{eff}^{\mathrm{int}}$ reflecting the trapping phenomenon for the polytrope spacetime having parameters from the trapping region in the $n$--$\sigma$ space. The effective potential is illustrated for the $n=3$ polytropes. Parameter $\sigma$ is evenly distributed between its maximum and minimum value. For each pair $\{n, \sigma\}$, loci of the stable and unstable circular null geodesics are marked. The $r_\mathrm{in}$ radius is constructed for $\sigma = \sigma_\mathrm{max}$ effective potential.   \label{fig8}}
\end{figure}

The effective potential $V{}^\mathrm{int}_\mathrm{eff}(\xi;n,\sigma)$ is constructed for the trapping polytropes with index $n = 3$ for various values of relativistic parameter from the interval $\sigma_\mathrm{min}(n) < \sigma < \sigma_\mathrm{max}(n)$ and illustrated in Fig.~\ref{fig8}. The stable (unstable) null geodesics correspond to the local maxima (minima) of $V{}_\mathrm{eff}^\mathrm{int}$. Their positions are given by values of $\xi$ close to 1 for the whole region of trapping. This means that for large configurations the trapping zones are located nearby the centre, but for small configurations they are located near the surface. The exact ratio $r_\mathrm{c}/R$ is obtainable using Tab.~\ref{tab1} for selected values of $n$ (for $\sigma_\mathrm{max}$).

We demonstrate the dependence of the extension of the trapping zone on the parameter $\sigma$. The zone extension is given by the intersection of the line $V_\mathrm{eff}(r = r_\mathrm{b}(n,\sigma)) = \mathrm{const}$ with the effective potential at $r < r_\mathrm{c}(n,\sigma)$, denoted as $r_\mathrm{in}(n,\sigma)$. The trapped null geodesics are restricted just to the region $r_\mathrm{in}(n,\sigma) < r < r_\mathrm{b}(n,\sigma)$.

\subsection{Local compactness radial profiles of the trapping polytropes}
Finally, we consider the behavior of the compactness of the trapping polytropic spheres. We present the compactness of the total polytropic spheres and the compactness related to their interior at $r_\mathrm{c}$, $r_\mathrm{b}$, and $r_\mathrm{extr}$ for the polytropes with maximal relativistic parameter, and for the central density $\rho_\mathrm{c} = 5 \times 10^{15}$~g\,cm$^{-3}$ in Tab.~\ref{tab2}. We can see that the compactness of the complete polytropes is very small, especially for $n$ close to $n=4$. However, surprisingly, even the compactness inside the polytrope at the centre of the trapping zone ($r_\mathrm{c}$) and its outer boundary ($r_\mathrm{b}$) is not close to the critical value of $C = 1/3$. There is always $C(n,\sigma_\mathrm{max}) < C\left(r_\mathrm{c}(n,\sigma_\mathrm{max})\right) < C\left(r_\mathrm{b}(n,\sigma_\mathrm{max})\right)$.

\begin{table}
\caption{List of compactness parameters describing polytropic fluid sphere of given $n$ having $\sigma=\sigma_\mathrm{max}$. Compactness $C(R)$, $C(r_\mathrm{c})$ and $C(r_\mathrm{b})$ are independent on central density $\varrho_\mathrm{c}$. Values of $C(r_\mathrm{extr})$ are calculated for central density $\varrho_\mathrm{c}=5\times 10^{15}$~g\,cm$^{-3}$. \label{tab2}}
\begin{ruledtabular}
\begin{tabular}{ccccccccccccccc}
 $n$ & $\sigma_\mathrm{max}$ & $C(R)$ & $C(r_\mathrm{c})$ & $C(r_\mathrm{b})$ & $C(r_\mathrm{extr})$\\ \hline
 4.0 & 0.80000 & $9.88\times 10^{-6}$ & 0.25492 & 0.29788 & 0.29794 \\
 3.9 & 0.79592 & $4.08\times 10^{-5}$ & 0.25595 & 0.29829 & 0.29834 \\
 3.8 & 0.79167 & $2.55\times 10^{-5}$ & 0.25703 & 0.29872 & 0.29874 \\
 3.7 & 0.78723 & 0.00124 & 0.25817 & 0.29914 & 0.29915 \\
 3.6 & 0.78261 & 0.00297 & 0.25937 & 0.29956 & 0.29956 \\
 3.5 & 0.77778 & 0.00454 & 0.26065 & 0.29998 & 0.29998 \\
 3.4 & 0.77273 & 0.00577 & 0.26199 & 0.30040 & 0.30039 \\
 3.3 & 0.76744 & 0.00660 & 0.26342 & 0.30081 & 0.30080 \\
 3.2 & 0.76190 & 0.00711 & 0.26493 & 0.30121 & 0.30120 \\
 3.1 & 0.75610 & 0.00768 & 0.26655 & 0.30160 & 0.30160 \\
 3.0 & 0.75000 & 0.00940 & 0.26828 & 0.30197 & 0.30198 \\
 2.9 & 0.74359 & 0.01432 & 0.27014 & 0.30230 & 0.30233 \\
 2.8 & 0.73684 & 0.02393 & 0.27215 & 0.30260 & 0.30266 \\
 2.7 & 0.72973 & 0.03783 & 0.27434 & 0.30283 & 0.30295 \\
 2.6 & 0.72222 & 0.05463 & 0.27673 & 0.30299 & 0.30318 \\
 2.5 & 0.71429 & 0.07308 & 0.27939 & 0.30301 & 0.30334 \\
 2.4 & 0.70588 & 0.09226 & 0.28241 & 0.30283 & 0.30340 \\
 2.3 & 0.69697 & 0.11152 & 0.28595 & 0.30228 & 0.30331 \\
 2.2 & 0.68750 & 0.13043 & 0.29057 & 0.30084 & 0.30304 \\
\end{tabular}
\end{ruledtabular}
\end{table}

\begin{figure}
\centering\includegraphics[width=0.99\linewidth,keepaspectratio=true]{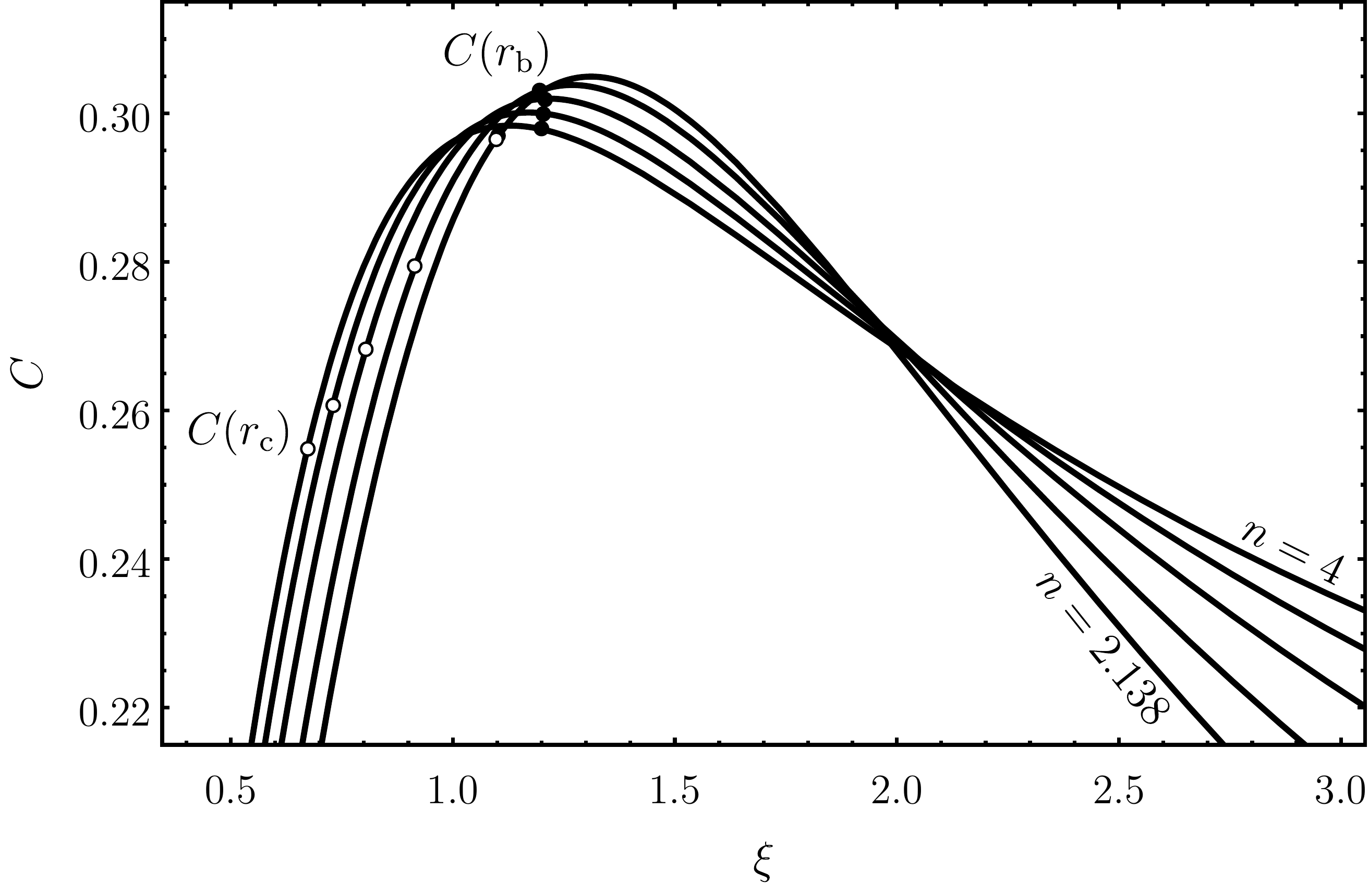}%
\caption{The radial profiles of the local compactness in the region of trapping and in its vicinity. Profiles are plotted for $n\in\{2.138,2.5,3,3.5,4\}$ and the corresponding maximal value of $\sigma$. On each  of the local compactness profile, loci of the stable and unstable circular null geodesics are marked.    \label{fig9}}
\end{figure}

One could intuitively expected that the compactness inside the region where trapping effect occurs is high and maximal nearby the radius $r_\mathrm{c}$. However, this is not true, as seen in Fig.~\ref{fig9} giving the dimensionless radial profiles of the compactness function $C(\xi)$ for selected values of $n$ and related maximal value of $\sigma$.

It is explicitly demonstrated that the maximal values of the compactness function $C(\xi)$ occur nearby (slightly above) the outer edge of the trapping zone and never cross the critical value of $C=1/3$. The maximal value of the compactness function increases with decreasing value of the polytropic index $n$. It seems that the trapping phenomenon is ruled by the strong gradient of the compactness function $C(r)$ rather than by the compactness itself.

\section{Conclusions}
In our study we demonstrate existence of standard general relativistic polytropes containing zone of trapped null geodesics. The trapping polytropes can exist, if the polytropic index $n \geq 2.1378$, and the relativistic parameter $\sigma$ is sufficiently high, but lower than the maximal value given by the causality limit. The critical value of the relativistic parameter related to the $n = 2.1378$ polytrope reads $\sigma_\mathrm{max}=0.681$. For whole range of polytropic indexes, the trapping zone can not exist, if $\sigma < 0.677$.

Estimates on limits on the polytropic indexes are presented for some equations of state in~\cite{2009RLO-CPP}. The possibility to apply different polytropic equations of state at different regions of the neutron star radial profile is mentioned in \cite{2009RLO-CPP}. Moreover, our preliminary searches indicate existence of the trapping zones in neutron star models related to sufficiently realistic equations of state.

The trapping polytropes do not fulfill the standard requirement on the existence of extremely compact objects stating that the surface has to be located under the photon circular orbit of the external spacetime, and $C > 1/3$, established for the configurations with uniformly distributed energy density \cite{2012SHU-NTE}. We have demonstrated inverse --- the compactness parameter can be much lower than the critical value of $C=1/3$. Moreover, even the local compactness radial profile $C(\xi)$ is not reaching this critical value, and its maximum lies outside the trapping zone. Seemingly, the gradient of the local compactness functions is decisive for occurrence of the trapping effect.

We have considered, if the existence of trapping polytropes could be physically relevant, namely, in the case of neutron stars. We thus assumed applicability of the polytropic equation of state up to the region where the gravitational mass of the polytropic configuration reaches the value of $M = m(r_\mathrm{extr}) = 2M_{\odot}$, given by the recent observational restrictions on the neutron star mass. The trapping polytropes representing neutron stars can exist, if the trapping zone is located under the radius of applicability $r_\mathrm{extr}$. We have shown that the trapping $n = 3$ polytropes can exist, if the central energy density reaches the value of $\rho_\mathrm{c} = 5 \times 10^{15}$~g\,cm$^{-3}$.

The trapping zones of the polytropes with index $n = 3$ (or $n \sim 3$) could be expected to give astrophysically relevant illustration of the effect of trapped null geodesics, as the $n = 3$ polytropes corresponds to the ultrarelativistic degenerated Fermi gas that could serve as an astrophysically relevant basic approximation of matter in the central parts of neutron stars \cite{1983ST-BHW}. Moreover, it is known that the realistic equations of state could be, at least partially, approximated by the polytropic equations of state \cite{2009RLO-CPP,2009OP-RNE,2010SLB-ESO}.

We expect that the trapping zones of the general relativistic polytropes could be relevant in the trapping of neutrinos and related cooling of neutron stars, or in the case of trapping of gravitational waves.

Stability of the trapping polytropes will be studied in a forthcoming paper. Of course, it could be also interesting to study possibility of existence of trapping zones in polytropes governed by alternative gravitational theories and, especially, in neutron stars governed by recently considered realistic equations of state.

\begin{acknowledgments}
ZS has been supported by the Albert Einstein Centre for Gravitation and Astrophysics financed by the Czech Science Agency Grant No. 14-37086G. J.N. and Z.S. were supported by the Silesian University at Opava internal grant SGS/14/2016.
\end{acknowledgments}


\bibliography{Polytropes_with_trapped_null_geodesics}

\end{document}